\documentclass[manuscript]{aastex}

\shorttitle{MIR View of W51A}
\shortauthors{Barbosa et al.}

\begin{document}


\title{A Mid Infrared View of the High Mass Star Formation Region W51A}


\author{C. L. Barbosa}
\affil{Laborat\'orio Nacional de Astrof\'{\i}sica, R. dos Estados Unidos, Bairro das Na\c c\~oes, CEP 37504-364, Itajub\'a - MG, Brazil}
\affil{Centro Universit\'ario da FEI, Departamento de F\'{\i}sica, Av. Humberto de A. C. Branco 3972, CEP 09850-901, S\~ao Bernardo do Campo - SP, Brazil}
\email{cassio.barbosa@pq.cnpq.br}

\author{R.D. Blum}
\affil{National Optical Astronomy Observatory, Tucson, AZ 85719}
\email{rblum@noao.edu}

\author{A. Damineli}
\affil{Instituto de Astronomia, Geof\'{\i}sica e Ci\^encias Atmosf\'ericas, Universidade de S\~ao Paulo, R. do Mat\~ao, 1226, Cid. Universit\'aria, S\~ao Paulo 05508-900, Brazil}
\email{augusto.damineli@iag.usp.br}

\author{P. S. Conti}
\affil{JILA, University of Colorado, Boulder, CO 80309-0440, USA}
\email{pconti@jila.colorado.edu}

\author{D. M. Gusm\~ao}
\affil{IP\&D -- Universidade do Vale do Para\'{\i}ba, Av. Shishima Hifumi, 2911. S\~ao Jos\'e dos Campos, SP, 12244-000, Brazil}
\email{danilo@univap.br}


\begin{abstract}
In this paper we present the results of a mid infrared study of G49.5-0.4, or W51A, part of the massive starbirth complex W51. Combining public data from the $Spitzer$ IRAC camera, and Gemini mid infrared camera T-ReCS at 7.73, 9.69, 12.33 and 24.56 \micron, with spatial resolution of $\sim$0.5\arcsec, we have identified the mid infrared counterparts of 8 ultracompact HII regions, showing that two radio sources are deeply embedded in molecular clouds and another is a cloud of ionized gas. From the T-ReCS data we have unveiled the central core of W51 region, revealing massive young stellar candidates. We modeled the spectral energy distribution of the detected sources suggesting the embedded objects are sources with spectral types ranging from B3 to O5, but the majority of the fits indicate stellar objects with B1 spectral types. We also present an extinction map of IRS~2, showing that a region with lower extinction corresponds to the region where a proposed jet of gas has impacted the foreground cloud. From this map, we also derived the total extinction towards the enigmatic source IRS~2E, which amounts to $\sim$60 magnitudes in the $V$ band. We calculated the color temperature due to thermal emission of the circumstellar dust of the detected sources; the temperatures are in the interval of $\sim$100 -- 150 K, which corresponds to the emission of dust located at 0.1 pc from the central source. Finally, we show a possible mid infrared counterpart of a detected source at mm wavelengths that was found by \cite{zap08,zap09} to be a massive young stellar object undergoing a high accretion rate.
\end{abstract}





\section{Introduction}

The W51 giant molecular cloud (GMC) is a complex of  compact radio sources and luminous infrared regions that are spread on the sky over an area of one square degree \citep{kumar04} near the Carina-Sagittarius arm tangential point ($l\sim49\fdg5, b\sim0\fdg2$). Two of these radio sources (W51A and W51B) are in fact complexes of HII regions and so, they are bright near infrared (NIR) sources. W51A harbors two radio continuum sources: G49.4 -- 0.3 and G49.5 -- 0.4 (G49.4 and G49.5 for short). Both sources were observed at radio wavelengths in great detail (maximum resolution of 3\arcsec) by \citet {meh94} and a total of 16 and six compact sources were found respectively in G49.4 and G49.5. The former is also the most luminous NIR source within W51, and one of the most luminous regions in the Galaxy. Low resolution infrared maps at 2 and 20 \micron\ show that G49.5 has two emissions peaks, called IRS~1 and IRS~2 that correlate quite well with the compact HII region sources W51e and W51d respectively \citep{wynn74,gww94}. Four (independent) clusters of masers can be identified in W51A: W51~NORTH, SOUTH (with two subclusters within SOUTH) and MAIN \citep[e.g.][]{genzel78}.  W51~MAIN and NORTH are also associated with the infrared peaks IRS~1 (that harbors the radio source W51e) and IRS~2 (that harbors the radio source W51d), respectively \citep{genzel82}. W51 may represent one of the most violent event of star formation in our Galaxy.

Among all the regions in this complex, G49.4 and G49.5 are the ones where star formation is most active. They are the youngest regions in W51, with estimated ages of $\sim$0.8 Myr and consequently, subject to a large average foreground extinction of $A_{V}\geqslant 9$ mag \citep{oku00}. Nevertheless, combining the luminosity and NIR colors of the detected objects and the radio data presented by \cite{meh94}, \citet{oku00} derived a population of 21 O stars within W51A, including those embedded massive young stellar objects (MYSO) that are ionizing ultracompact HII (UCHII) regions: 3 O9, 2 O8, 1 O7, 7 O6 4 O5 and 4 O4 stars. This large population of massive stars is reflected in a excess of objects more massive than 30 $M_{\sun}$, evident on the Initial Mass Function of the cluster.

Different distances have been found for W51A  in the literature. The kinematic distance to the whole complex is 5.5 kpc \citep{kolp03}, in contrast with \citet{lys08} who derived a distance of 2.2 kpc by means of spectroscopic parallaxes of hot stars toward W51A. \citet{xu09} obtained a distance of 5.1 kpc measuring the trigonometric parallaxes of maser sources in W51~NORTH (W51~IRS~2). More recently, \cite{sato10} derived a distance of 5.4 kpc for W51~MAIN/SOUTH, also by means of trigonometric parallax of water masers. It appears likely the distance derived by \cite{lys08} may not apply to W51, but instead to an intervening stellar population along the line of sight. This is consistent with the fact that the stars observed by those authors are spread in a much larger region and with much smaller interstellar reddening. In the present work, we adopt 5.5 kpc, quoted by \cite{kolp03} as the distance to W51, which is in agreement with other measurements targeting the G49.480-0.386 region.

W51~IRS~2 was observed at mid infrared (MIR) wavelengths by \citet{oka01} and \citet{krae01}. Okamoto and collaborators presented images taken through narrow band filters ($\Delta\lambda\sim0.23$ \micron) between 8.0 and 13.2 \micron, including filters for the fine structure lines
[NeII] (12.1 \micron), [ArIII] (8.9 \micron) and [SIV] (10.5 \micron) and low resolution spectra (R$\sim$100) of 7 compact sources identified
from their maps. Matching the radio continuum map of \citet{gaume93} and their 12 \micron\ map, \citet{oka01} were able to correlate the MIR peak IRS~2W to the main radio source of this region, the UCHII region W51d. The (NIR) counterpart and hence the ionizing source of
this UCHII region was identified by \citet{lys08} matching high resolution  maps taken at radio and MIR wavelengths with a high resolution
image taken at the $K$ band. \citet{barb08} classified this ionizing source as an O3 star, based on a medium resolution spectrum (R$\sim$5200) in the $K$ band. \cite{krae01} using a set of narrowband filters were able to derive the color temperature, the mass of hot dust and column density of seven sources in W51~IRS~2, including IRS~2E (their source KJD~3) detecting the deep silicate absorption profile in this source, speculating that it could be the signature of a disk surrounding a MYSO.

In this paper, we present the results of our MIR observations as part of a campaign aimed to study W51A. The paper is organized as follows: in section \ref{data} we present the details of the data used in this work; in section \ref{results} we present and discuss the results obtained after processing the data; in section \ref{summ} we present the summary and our conclusions.

\section{The data} \label{data}

In this paper we used two sets of MIR data, one was from the {\it Spitzer} public database and used mainly to correlate MYSOs with UCHII regions. The other    set of data was obtained at the Gemini South observatory. We give the details of both sets below.

\subsection{Spitzer data}

Data were obtained from the {\it Spitzer} Heritage Archive as part of the GLIMPSE survey legacy program \citep{benj03}. We retrieved the images from the level 2 processed data, which is also known as post basic calibration data. They are large ($\sim$2\fdg 6$\times$0\fdg 3) flux calibrated mosaics built from individual (5\farcm2$\times$5\farcm2) images from the Infrared Array Camera \citep[IRAC]{fazio04}. We used the four mosaics (Program: 187; Campaign ID: 751; AORKEY: 12241920, PI: Ed Churchwell) taken with all four IRAC filters centered at 3.6, 4.5, 5.8 and 8.0 \micron\ to locate  W51~IRS~1 and IRS~2 in the field, the angular resolution of the mosaics is 1.2\arcsec /pixel.

\begin{figure}
\epsscale{2.0}
\plottwo{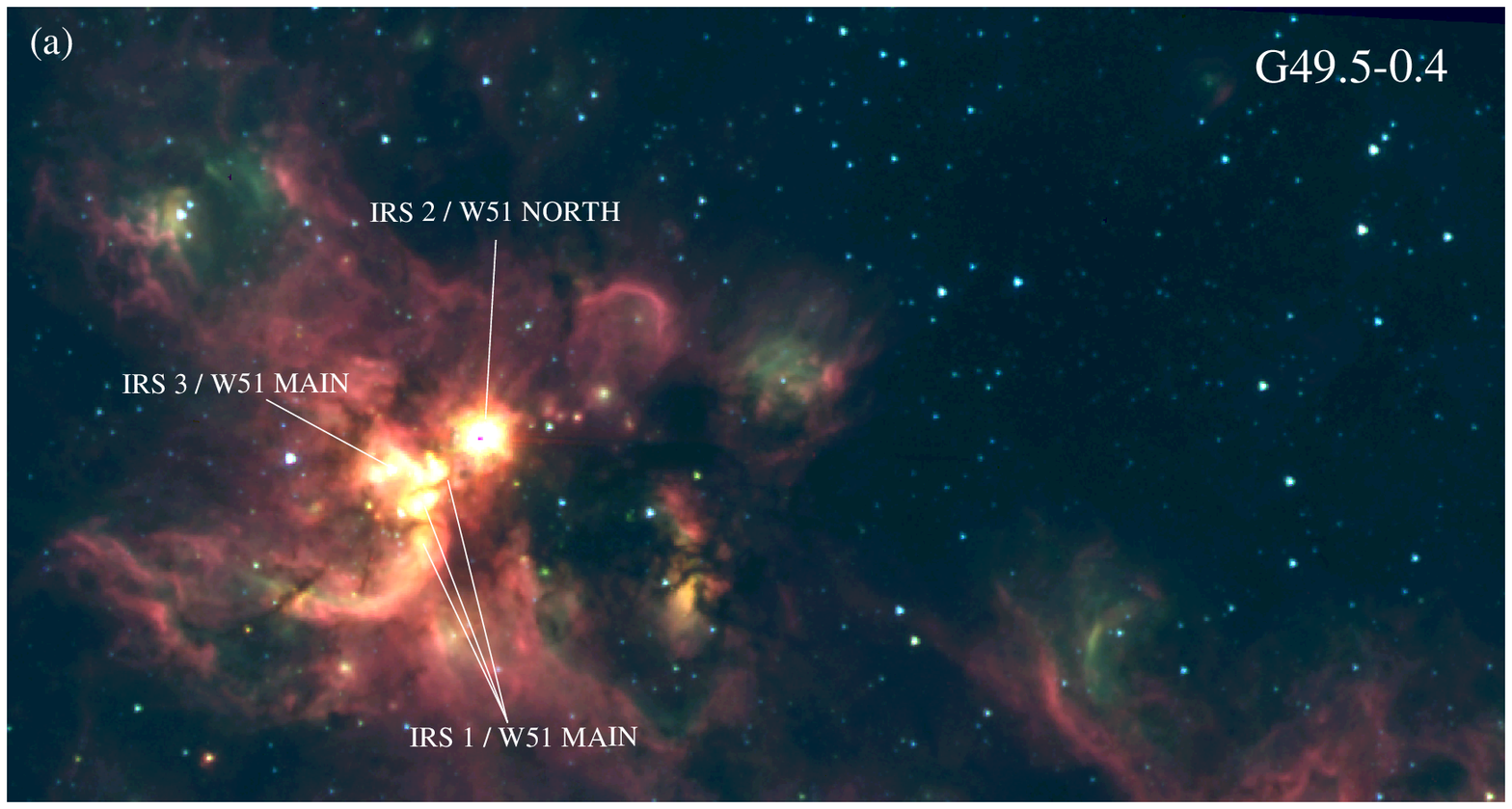}{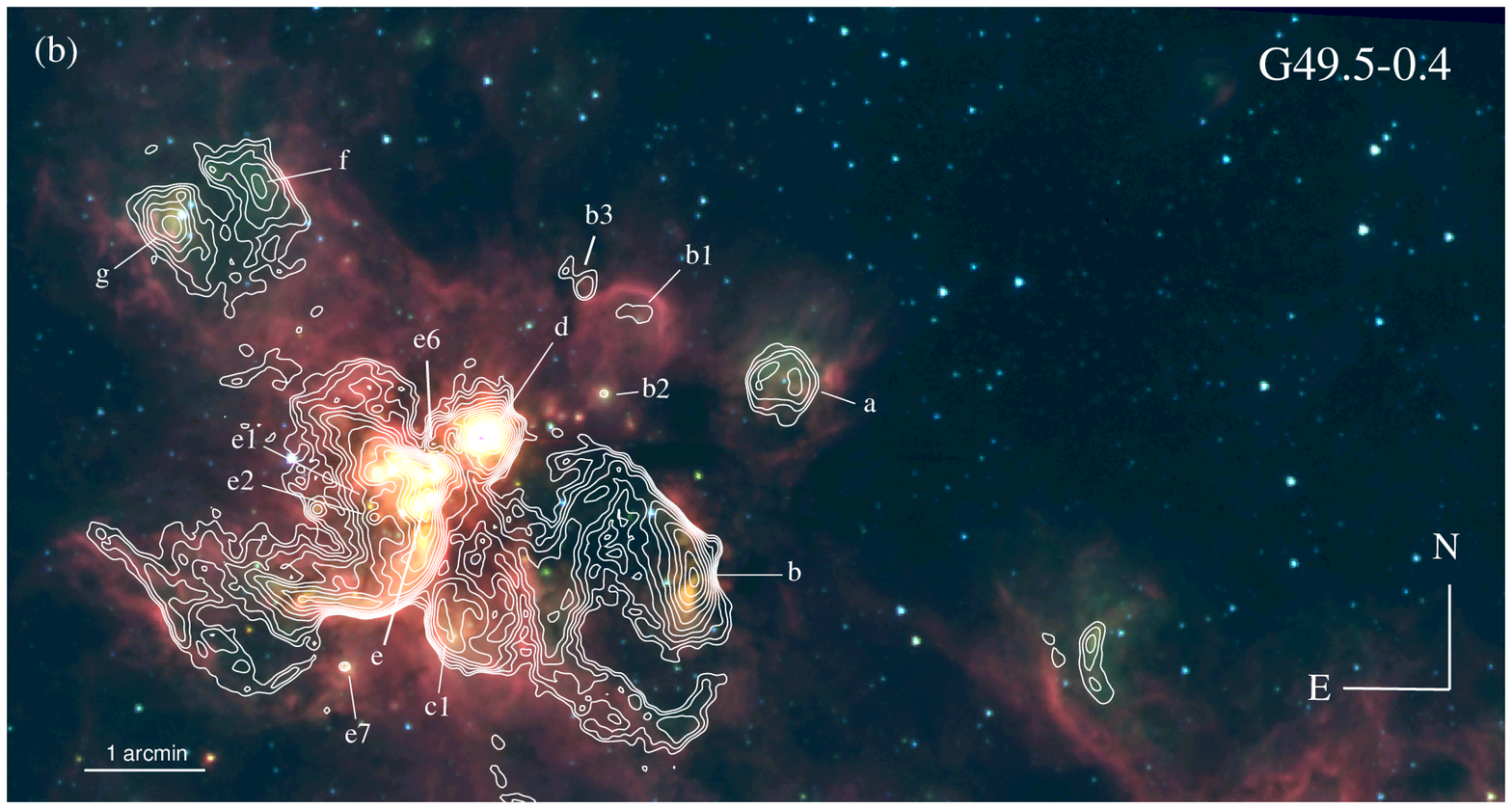}
\caption{Panel (a): color composite image of G49.5-0.4 constructed from three IRAC filters: red represents the 5.8 \micron\ filter, green is 4.5 \micron\ and blue is 3.6 \micron. Infrared sources IRS~1 and IRS~2, from \citet{wynn74}, and IRS~3 from \citet{gww94} are indicated. Panel (b): the same composite image, but with radio emission curves overplotted in white. The level curves represent the emission at 6 cm, the beam width is 5\farcs$2\times$4\farcs5 (PA=89\degr). The levels range from 0.01 to 1.8 Jy beam$^{-1}$ logarithmically spaced in 15 steps. Labels represent UCHII regions identified by \citet{meh94}. For reference, the coordinates of b$_{2}$ are: $\alpha$ = 19:23:35.85; $\delta$ = 14:31:27.39 (J2000). The match between the MIR and radio images is good within 2\arcsec.}
\label{irac_rad}
\end{figure}

In Figure \ref{irac_rad} panel (a) we present a mosaic of IRAC images composed by three of its filters, color coded as 5.8 \micron\ as red, 4.5 \micron\ as green and 3.6 \micron\ as blue. Sources IRS~1 and IRS~2 identified by \cite{wynn74} and IRS~3 identified by \cite{gww94} are indicated. The image taken through the fourth filter centered at 8.0 \micron\ could not be used as IRS~1 and IRS~2 are  saturated. Panel (b) of Figure \ref{irac_rad} shows the same color composite image over which we plotted the free-free radio emission at $\lambda=6$ cm as level curves. These level curves were obtained from the electronic version of plate 56 presented by \cite{meh94}.

\subsection{Gemini data}\label{gemdata}

The second part of the data were obtained at the Gemini South observatory (program: GN-2006-Q-17; PI: C\'assio Barbosa), using the Thermal-Region Camera Spectrograph (T-ReCS), which is fully described by \cite{tel98}. The instrument contains a Si:As Raytheon SBRC detector, with 320$\times$240 pixels of 50 \micron, each. When mounted on the telescope it gives a field of view (FOV) of 28\farcs8$\times$21\farcs6, with a plate scale of 0.09\arcsec/pixel.

\begin{figure}
\centerline{\rotatebox{270}{\scalebox{0.5}{\includegraphics{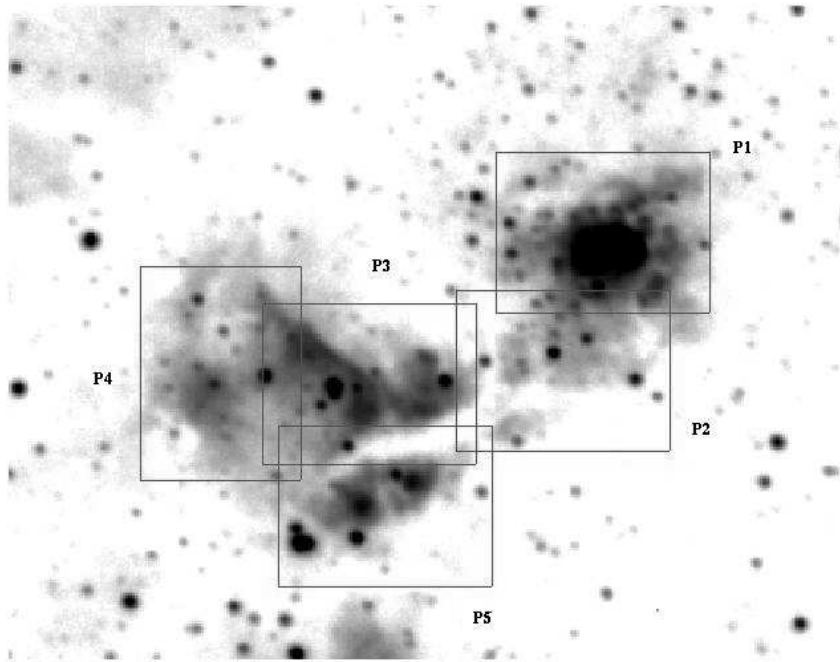}}}}
\caption{Finding chart for the studied regions. Each rectangle corresponds to the FOV of the T-ReCS camera (28\farcs8$\times$21\farcs6). P$_{3}$, P$_{4}$ and P$_{5}$ partially cover IRS~1. P$_{2}$ covers the region between IRS~1 and IRS~2 and P$_{1}$ entirely covers IRS~2. North is up and East is to the left.}
\label{mosaic}
\end{figure}

The data were taken pointing the telescope at five different positions, each pointing corresponding to a single FOV with a small overlap among them. Each position was imaged through four narrowband filters. Figure \ref{mosaic} is a $K-$band image from 2MASS \citep{skru06} in which we show the positions of the five pointings as the rectangles representing one FOV of the camera. In Table \ref{journal} we present the details of the observations, which were performed on four nights in May 2006: 07, 08, 14 and 15.

The data were obtained in a standard chop-and-nod procedure to subtract the thermal background. We chose the maximum possible chopping throw (15\arcsec)  toward directions with evidence of low background emission, after inspecting both 2MASS $K$ band and {\it Spitzer} images at 8.0 \micron. Even chopping 15\arcsec\ away from the target, T-ReCS chops on chip, which produces artifacts corresponding to the negative beam in the background subtracted image. We set 200 seconds as exposure time on source for all filters.

The data were reduced under IRAF\footnote{IRAF is distributed by the National Optical Astronomy Observatories} environment using the scripts for MIR data reduction available in the Gemini IRAF package. The reduction process consists basically of stacking the multiple images obtained chopping and nodding the telescope then subtracting the sky beam. All images were flux calibrated using the set of standard stars observed as part of the baseline calibration program. The standards were observed at similar airmass as the targets. The uncertainties regarding airmass correction and the fluxes of the standards themselves limit the accuracy to about 10\% in the measured fluxes.

We extracted the fluxes of the resolved sources through circular aperture photometry, but an important caveat must be outlined regarding the local background. Although we carefully inspected public available images to define a suitable sky position to subtract the thermal background emission, the targets themselves are surrounded by extended emission, as W51 is an active young star birth site. Given the limited chopping throw of the secondary mirror (15\arcsec) we were unable to observe distant positions to use them as sky in order to subtract from the target positions. To overcome this difficulty we inspected all calibrated images measuring the local background emission. We assumed the lowest emission found in the images as the background to be subtracted from the target fluxes. This approach was performed by \cite{barb03} in a similar study of the massive star birth site NGC 3576 with good results.

\section{Results}\label{results}
\subsection{Identifying the UCHII regions}

Massive young stars are noticeable radio sources in early stages of their lives \citep{church02}. They can be detected as UCHII regions representing the moment when the accretion process is finished or is greatly reduced. High spatial resolution and high sensitivity observations at radio wavelengths \citep[e.g.][]{gaume93,meh94} of IRS~1 and IRS~2 show a plethora of compact sources. The number of emitted Lyman continuum photons obtained from the analysis of their free-free emission (1.3 $\leq\lambda\leq$ 6 cm) suggests that these UCHII sources are embedded young massive stars.
 
Panel (a) of Figure \ref{irac_rad} reveals complex structures within G49.5 -- 0.4, representing the interaction between MYSOs and their natal clouds of dust and gas. Sources IRS~1, IRS~2 and IRS~3 are indicated. The bluish objects are those subject to a low interstellar reddening, presumably foreground stars. The green extended emission permeating the clouds is often found in massive star birth sites when the IRAC 4.5 \micron\ channel is coded as green in color composite images \citep[e.g.][]{cyg08}, and they are called ``green fuzzies", or ``extended green objects" (EGO). This effect is sometimes associated with outflows, shocks and atomic/molecular hydrogen emission. An attempt to establish the spectral carrier of this enhancement in the green channel was conducted by \cite{deb10}, as emission lines from Br$\alpha$ ($\lambda$ = 4.05 \micron), H$_{2}$ ($\lambda$ = 4.69 \micron) and [FeII] ($\lambda$ = 4.11, 4.43, 4.61 4.67 or 4.89 \micron) are expected to be observed within the bandwidth of the 4.5 \micron\ filter. Although this study is not conclusive in identifying the origin(s) of the green extended emission, all lines listed above are often observed in birth sites of massive stars, therefore any of these lines can be associated with the presence of MYSOs.  

Panel (b) of Figure \ref{irac_rad} shows the same MIR image, but with level curves from free-free emission at $\lambda=$ 6 cm overplotted. The labels represent the UCHII regions observed by \cite{meh94} in high spatial resolution. The composite image in panel (b) is useful to reveal the sources of radio emission in the MIR, in other words, we can pinpoint the MIR counterparts of the UCHII regions. The extended MIR emission follows the radio emission, as the gas and dust of HII regions are mixed, and one can be traced by the other. More interesting is the fact that the green extended emission discussed before follows the radio emission even closer, e.g. the UCHII region W51a. This is a shell like UCHII region \citep[in the scenario proposed by][]{wc89a} which appears to be filled by the green extended emission. This fact is one of the arguments favoring Br $\alpha$ as the main responsible for the quoted green emission.

Mid infrared counterparts for the UCHII regions a, b$_{1}$, b$_{2}$, c$_{1}$, e$_{1}$, e$_{2}$, e$_{7}$ and g can be readily identified. Source b$_{3}$ does not show any MIR counterpart, however it can be seen that its emission flows from a dark molecular cloud, which indicates that the ionizing source of this UCHII region is still buried in its natal cloud. Source d is a bright MIR emission peak \citep[see][]{lys08} and e is associated with the central emission peak of IRS~1. We could not find any MIR counterpart for regions b$_{3}$ and f, even after inspecting the image at 8 \micron, taken by $Spitzer$. The morphology of the UCHII region f suggests that it may be just part of a bubble of gas ionized by region g. In fact, \cite{meh94} derived a similar spectral type for both ionizing objects: O6.5 for g and O6 for f.

\subsection{The high resolution images}\label{highres}

\begin{figure}
\centerline{\scalebox{0.7}{\includegraphics{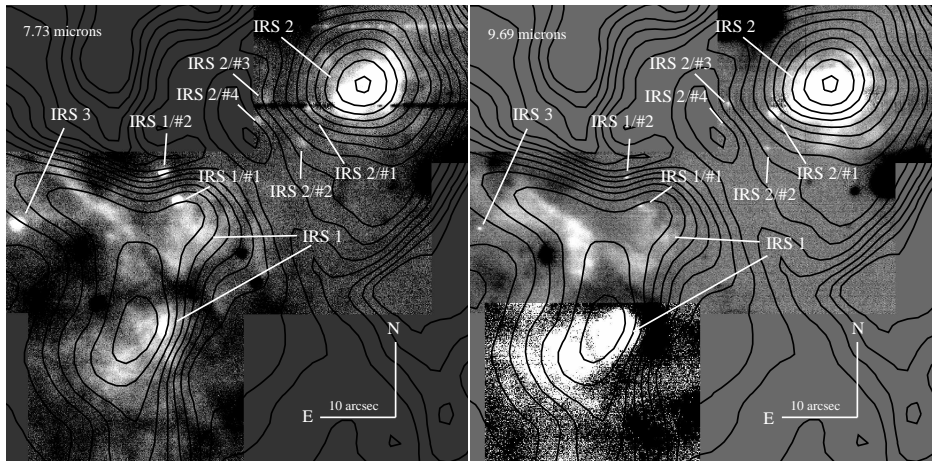}}}
\caption{Mosaics built from T-ReCS images. In each panel we labeled the sources detected in the mid infrared and plotted (in black) the same level curves or radio emission plotted on the IRAC image of Figure \ref{irac_rad}. The intensity was stretched to enhance weak objects like those around IRS~2. The darker regions in all mosaics represent the negative beam of the chopping procedure (see text), except the dark lane between the two components of IRS~1. This dark lane is detected also at shorter wavelengths. {\it Left:} Mosaic built from images observed with the Si-1 filter (7.73 \micron). {\it Right:} Mosaic built from images observed with the Si-3 filter (9.69 \micron). In this mosaic, the image corresponding to P$_{4}$ has a significantly lower signal to noise ratio.}
\label{trecs1}
\end{figure}

\begin{figure}
\centerline{\scalebox{0.7}{\includegraphics{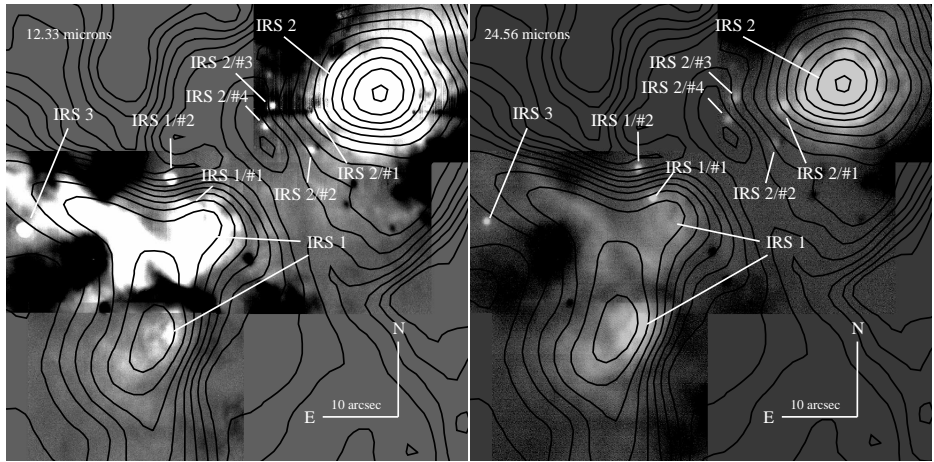}}}
\caption{Same as Figure \ref{trecs1}, but imaged with different filters. {\it Left:} Mosaic built from images observed with the Si-6 filter (12.33 \micron) In this mosaic, the image corresponding to P$_{4}$ has a significantly lower signal to noise ratio. {\it Right:} Mosaic built from images observed with the Qb filter (24.56 \micron).}
\label{trecs2}
\end{figure}

In this section we present the high resolution images taken with the mid-infrared camera T-ReCS. The flux calibrated images were combined in mosaics, following the pattern displayed in Figure \ref{mosaic}, except for P$_{4}$. We did not detect any point source in this field, besides the one that is also present in field P$_{3}$. Figure \ref{trecs1} shows the mosaics obtained with filters centered at 7.73 \micron\ (left) and 9.69 \micron (right). Figure \ref{trecs2} shows similar mosaics, but composed by images taken with filters centered at 12.33 \micron\ and 24.56 \micron. In both figures, we stretched the intensity level to reveal weak sources and because of that, sources within IRS~2 are not visible due to the circumstellar emission. We will return to IRS~2 with more details in the next section.

In both figures, IRS~1, IRS~2 and IRS~3 are evident. IRS~1 is composed of three emission peaks (only the two northern ones were observed) and does not show any pointlike source. IRS~2 is a well known compact HII region and IRS~3 is a resolved infrared source. Objects IRS~2/\#2 -- \#4 and IRS~1/\#1 and \#2 are bright sources in all images, except in the image taken with the filter centered at 9.69 \micron. This filter is centered very close to the center of the silicate dust absorption profile and that indicates that these sources are embedded in dense dust cocoons, which in turn, is evidence that these sources are quite young. Object IRS~2/\#1 and W51d$_{1}$ are exceptions: they do not show the absorption feature but W51d may have a hint of absorption. This may indicate that these three sources could have dissipated their dust cocoons, cleaning their local environment, but they are still viewed behind a molecular cloud.

The coincidence between the position of the radio emission and the MIR mosaics is good within $\sim$2\arcsec. The correlation in the position was based on the position of the UCHII region W51d and the MIR emission peak in IRS~2, following the same procedure of \cite{lys08}.  Comparing the composite image of Figure \ref{irac_rad} and the mosaics of Figures \ref{trecs1} and \ref{trecs2} we note that this positional match is actually a little better than 2\arcsec, and this value can be used as an upper limit. Figures \ref{trecs1} and \ref{trecs2} confirm, with more details, that the UCHII region W51e correlates with the middle peak of IRS~1. None of the detected objets in these mosaics can be associated with any of the UCHII regions labeled in Figure \ref{irac_rad}.

\begin{figure}
\centerline{\scalebox{0.5}{\includegraphics{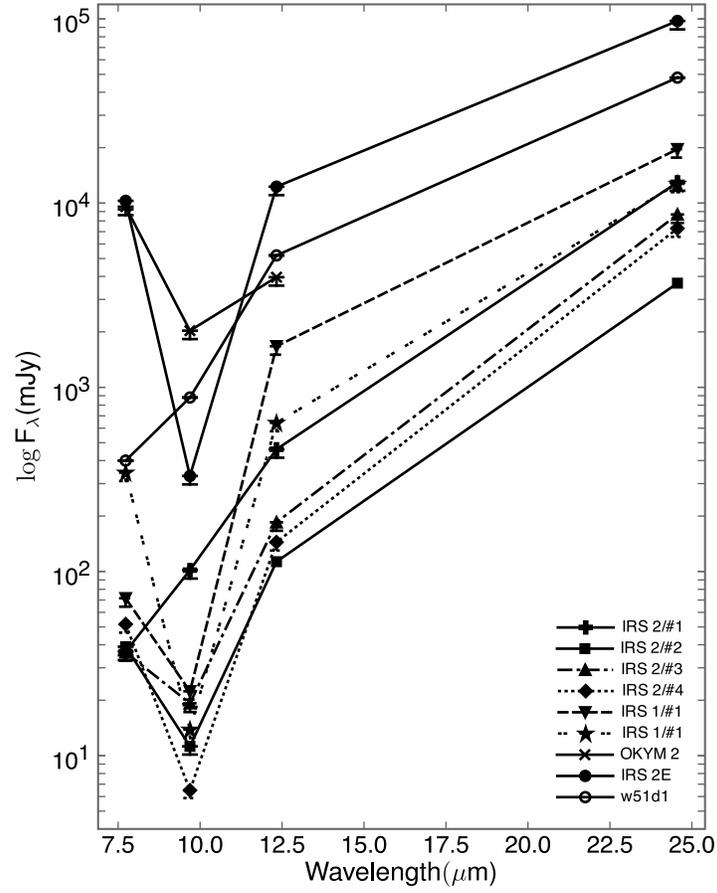}}}
\caption{Mid infrared fluxes extracted from the mosaics of Figures \ref{trecs1} and \ref{trecs2}. Due to the low S/N ratio, fluxes quoted for objects IRS~2/\#2, \#3 and \#4 must be taken as upper limits, specially at 7.73 and 9.69 \micron.}
\label{obj_flx}
\end{figure}

From the flux calibrated images showed in Figures \ref{trecs1} and \ref{trecs2} we extracted the fluxes of all objects through aperture photometry, as discussed in section \ref{gemdata}. The fluxes for all objects in each filter are presented in Table \ref{Av}. From the fluxes quoted in that table, we plotted the SED of each source in Figure \ref{obj_flx}. The SEDs of the sources presented in this figure are typical of MYSOs \citep[e.g.][]{barb03}, most show the silicate dust profile in absorption, which is indicative of the dusty environment in which they are embedded. The exceptions to this scenario (noted above) are sources IRS~2/\#1 and W51d$_{1}$, which show monotonic crescent fluxes towards longer wavelengths. Although such SEDs are observed in MYSOs also, the errors of the individual points in the SED do not leave room to a suggest any possible absorption in that filter.
 
\subsubsection{Color Temperature and Extinction}\label{colext}

We used the fluxes at two different wavelengths to derive the color temperature of the detected sources. Since the temperature determines the ratio of blackbody flux densities at any two wavelengths \citep{ball96} we can use the fluxes quoted in the Table \ref{Av} to calculate the color temperature of the sources; temperatures were obtained from the fluxes measured at 12.33 \micron\ and 24.56 \micron. At these two wavelengths, we estimate the color temperature of the dust from the following expression, where $R$ is the ratio $F_{24.56}/F_{12.33}$:

\[
T_{d} (\mbox{K})=580.4/[\ln{(7.93R)}]
\]

Likewise, we can derive an estimate of the total reddening for each detected source  estimating the silicate optical depth, using the fluxes quoted in Table \ref{Av}. Following the procedure described by \cite{gez98}, we obtained a ``pseudo-continuum" at the center of the silicate absorption feature ($\sim$ 9.8 \micron) interpolating the fluxes measured at 7.73 and 12.33 \micron. Ratioing this pseudo-continuum flux and the flux measured at 9.69 \micron, we can estimate the absorption due to the dust mixed to the circumstellar gas. This ratio represents the amount of extinction due to the dust ($A_{9.8}$), which can be converted to visual extinction assuming $A_{V}=16.6\times A_{9.8}$ \citep{rieke85}. The procedure adopted here does not account for the dust silicate emission that arises from inner, and therefore hotter, layers of dust. This underlying emission can be neglected for pointlike sources, but may be intense enough to ``fill in" the silicate absorption for more extended objects like the radio sources. This could explain the absence of this feature in the SEDs of sources IRS2/\#1, W51d$_{1}$ and W51d \citep[see][]{gill75}.

Although the visual extinctions quoted in the Table \ref{Av} for the weakest objects (IRS~2/\#2 and \#4) are somewhat uncertain due to the low S/N ratio of the measured flux of the sources, they are compatible with those expected for YSOs embedded in such environment. Table \ref{Td} shows, for each object, the dust color temperature, the extinction at 9.8 \micron\ and corresponding visual extinction.

The results presented in Table \ref{Td} are typical for young stellar objects in the early stages of formation \citep[e.g.,][]{ball96, barb03}. The temperatures obtained are in the range 96 K to 144 K. According to the dust emission models in circumstellar environments of \cite{wolf94}, these temperatures correspond to the thermal emission of dust located at $r\sim$ 0.1 pc from the object, considering only one central source of radiation, i.e., the object is considered in isolation, which may not be the case for IRS~2E, as this object is in dense cluster of MYSOs 
(one of which is a know early O-type star). We can compare the results for the temperature obtained for IRS~2E and W51d$_{1}$ with those obtained by \cite{krae01}. In their study IRS~2E is labeled KJD~3, and W51d$_{1}$ is labeled KJD~4, and $T_{d} = 172$ K for IRS~2E and $T_{d} = 145$ K for W51d$_{1}$, which are in good agreement with the corresponding results quoted in Table \ref{Td}, even though \cite{krae01} used the ratio of the fluxes measured at 12.5 and 20.5 \micron.

\subsubsection{The SED of the sources}\label{sed_sec}

\begin{figure}
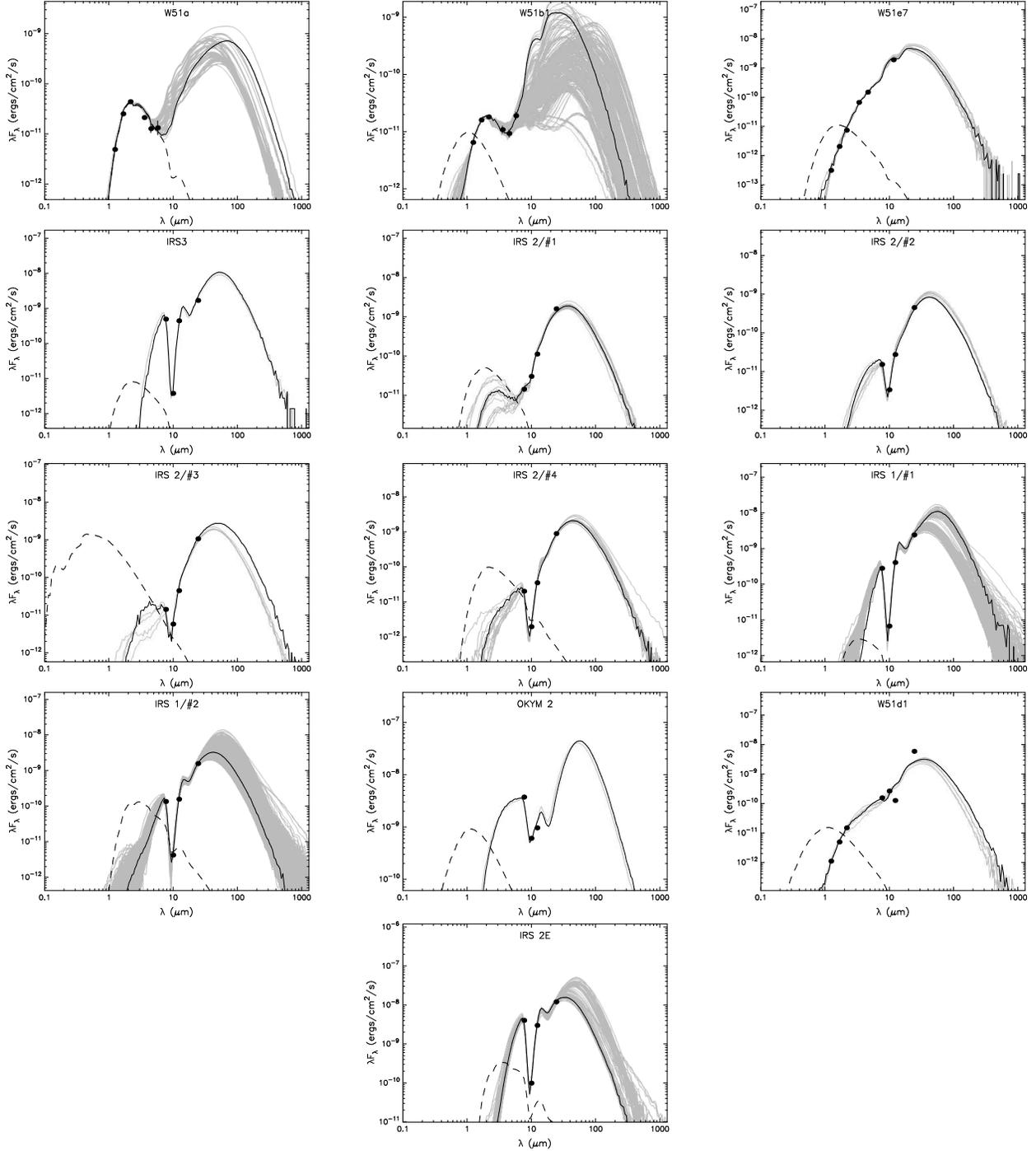

\scalebox{0.42}{\includegraphics{w51a.eps}}
\scalebox{0.42}{\includegraphics{w51b1.eps}}
\scalebox{0.42}{\includegraphics{w51e7.eps}}
\scalebox{0.42}{\includegraphics{irs3.eps}}
\scalebox{0.42}{\includegraphics{irs2-1.eps}}
\scalebox{0.42}{\includegraphics{irs2-2.eps}}
\scalebox{0.42}{\includegraphics{irs2-3.eps}}
\scalebox{0.42}{\includegraphics{irs2-4.eps}}
\scalebox{0.42}{\includegraphics{irs1-1.eps}}
\scalebox{0.42}{\includegraphics{irs1-2.eps}}
\scalebox{0.42}{\includegraphics{okym2.eps}}
\scalebox{0.42}{\includegraphics{w51d1.eps}}
\centerline{\scalebox{0.42}{\includegraphics{irs2e.eps}}}
\caption{Spectral energy distributions for detected objects within W51. In each panel, black circles represent fluxes extracted from the images and/or obtained in public databases, black lines are the best fit model for the data points, gray lines represent subsequent good fits for the observational data and dashed lines represent a stellar photosphere only affected by interstellar extinction.\label{sed_g1}}
\end{figure}

In order to investigate the nature of some of the detected sources in the
images, we used the fluxes extracted from the T-ReCS images and public
data from 2MASS and Spitzer projects (when available) to fit the SEDs
of the detected objects, using the SED fitting tool of \cite{robi07},
which, in turn, uses a grid of 2D radiative transfer
models. The fitting tool uses a grid of radiation transfer models of
20,000 YSO models, each one can be viewed through 10 different angles,
covering a range of stellar masses from 0.1 to 50 M$_{sun}$, from early
stages of envelope infall to the late passive disk-only stage. In the
former scenario, the radiation emitted for the central source is
entirely absorbed by the dust envelope, while in the later and more
evolved scenario, part of the radiation is allowed to escape through
bipolar cavities. The SEDs are fitted using a set of 14 parameters, but at a particular evolutionary stage, only a a few of them are importante. At earlier stages, the SEDs are most affected by: envelope accretion rate, the opening angle of the bipolar cavities, the inclination to the line of sight, the disk/envelope inner radius, the stellar temperature and the disk mass. The models do not include the contribution of hot dust. For further details see also \cite{robi06}, \cite{whit03b,whit03a}. The best fit models are presented in Figure \ref{sed_g1} and data derived from the models in Table \ref{dat_g1}.

The fitting tool does not return a unique model for the SED and, therefore, a unique set of parameters such as stellar age, stellar mass, stellar radius, stellar temperature, accretion rate, characterization of the envelope and the disk, total luminosity, ambient density and the circumstellar extinction. The spectral types derived from modeled SED were obtained according to the relationship between mass and spectral type listed by \citet{blum00}. For a few sources only, we can compare the spectral types obtained by the SED fitting tool with those in the literature, and the comparison is shown in the Table \ref{sp_comp}. The results shown in the tables above must be taken with care. Our results must be biased toward late O/early B stars, as the SED fitting procedure is limited to models with stellar mass less than 50 $M_{\sun}$, which precludes the identification of objects with spectral types earlier than O5.

The most discrepant result between estimates from the fitting tool and those found in the literature is the case for W51a. For this object, the SED suggests a B3 type star, while its radio emission suggests an O6 type star. We have no explanation for the discrepancy; extending the SED to longer wavelengths (i.e. new data) might clarify this situation.

Overall the results of the model fits are consistent with expectations. Most of the embedded objects are B type. The O stars have already emerged from their cocoons due to the much more energetic impact on their environment. The timescales for B types to push their birth material away is longer, and hence we are more likely to observe them in this phase. 

The stellar ages quoted in Table \ref{dat_g1} are obtained through an iterative process that randomly samples ages between $t_{min}=10^{3}$ yr and $t_{max}=10^{7}$ yr and compares them to the lifetime of a randomly sampled stellar mass between $M_{1}=0.1 M_{\sun}$ and $M_{2}=50 M_{\sun}$. This process continues until an adequate stellar age converges to that presented in the pre-main-sequence and main-sequence evolutionary tracks of \cite{bern96} for a given stellar mass. The evolutionary age of the central star is not a parameter in the radiation transfer code and is only used by the fitting tool to get a coherent radius and temperature of a suitable stellar model from evolutionary tracks of \cite{bern96} for M$_{\star}\geqslant 9$ M$_{\sun}$ and \cite{siess00} for M$_{\star}\leqslant$7 M$_{\sun}$, as well as approximate ranges of disk and envelope parameters. It is interesting to note that the quoted ages are less than $10^{5}$ in accordance to the expected lifetime of MYSOs.

\subsection{IRS~2}\label{irs2}

In this section we will present a more detailed study of IRS~2. This region is known to harbor a plethora of young stellar objects, many of them MYSO candidates \citep[e.g.,][]{gww94, oku00, oka01, lys08, barb08}. About a dozen OB candidates and at least 4 UCHII regions can be found in this compact HII region, however, until now only one hot star has been spectroscopically classified. \cite{barb08} identified the ionizing source of the UCHII region W51d as an O3 star. 

IRS~2 has two MIR peaks, named IRS~2E (on the eastern part of the cloud) and IRS~2W (on the western part of the cloud). The eastern peak is associated with a stellar object that becomes the brightest object in this region for $\lambda >$2 \micron.  The western peak, instead is not associated  with any point-like object, but is a patch of bright extended emission. \cite{lys08} show that this MIR peak is coincident with UCHII region W51d and the radio emission from this UCHII has a double peak. The most intense of the radio peaks corresponds to the O3 star identified by \cite{barb08}. The secondary radio peak is associated the MIR peak IRS~2W and also apparently with two point-like sources detected in new deeper high-resolution images taken in the near infrared (Barbosa et al., in prep.). We speculate that multiple MYSOs could be deeply embedded within IRS~2W (the coincidence with the second radio peak of W51d suggests they may be massive.

Although this region has been studied for decades, only recently could its young stellar population be resolved in detail \citep[e.g.][]{kumar04}. \cite{lacy07} using high resolution MIR spectroscopy found a massive jet emerging from a molecular cloud hitting a foreground HII region. Although they could not locate the source of this jet, their data, associated with a high-spatial resolution $K$-band image suggest that this jet has precessed by $\sim$90\degr\ or the gas could be have been emitted in a fan-like structure. Lacy and collaborators speculate that the jet is part of the same outflow that forms the high velocity H$_{2}$O and SiO masers peaks located at $\sim$ 1\arcsec\ to the south of the position where the high velocity gas impacts the HII region. At the position of the maser peaks, \citet{zap08} found a MYSO, possibly an early O type star with 40 M$_{\sun}$ still in formation. Moreover, \cite{zap09} found evidence of a process of gas accretion through a molecular ring and also a massive (200 M$_{\sun}$) and highly collimated ($\sim$14\degr) bipolar outflow. It is beyond the scope of the this paper to investigate the nature of this jet, however both the position of the source and the orientation of the outflow suggest that they may be connected.

\begin{figure}
\centerline{\scalebox{0.9}{\includegraphics{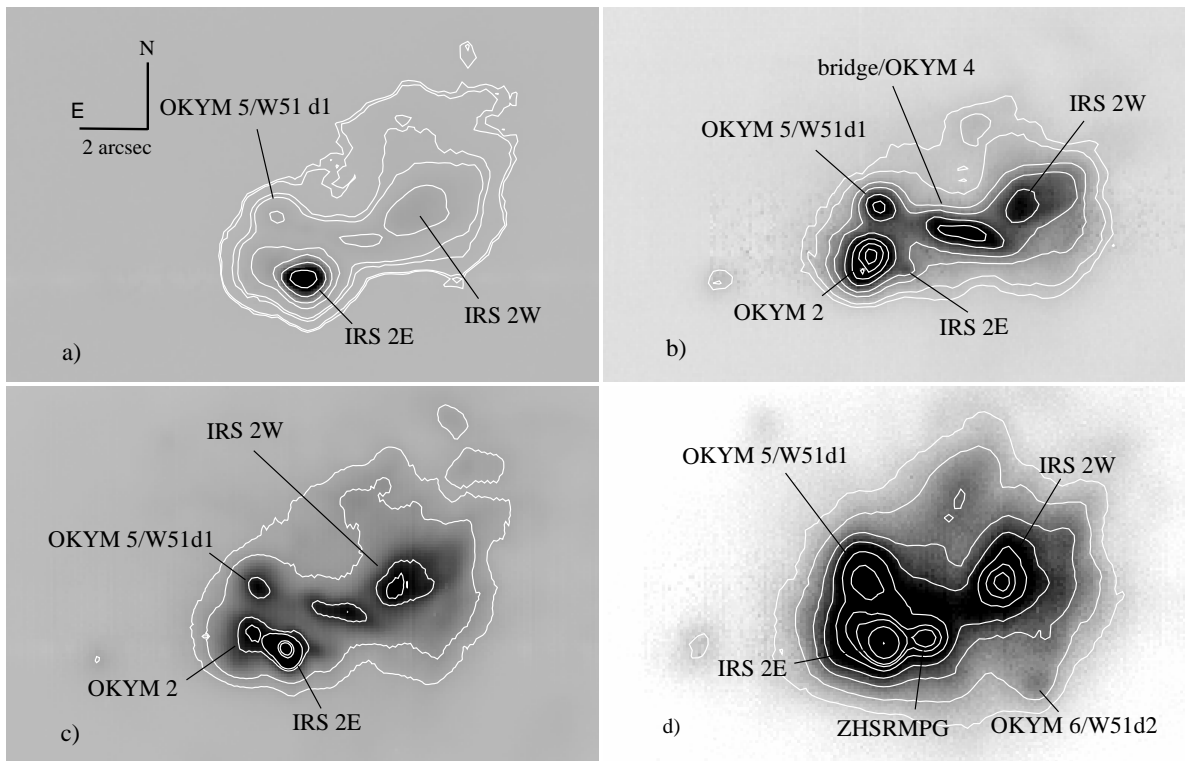}}}
\caption{IRS~2 seen through the filters used in this work. The white level curves indicates the MIR emission and are in arbitrary units, just to outline the sources and the extended structures found in this region.  (a) 7.73 \micron; (b) 9.69 \micron; (c) 12.33 \micron; (d) 24.56 \micron.  In panel (b) we identify the extended structure labeled as ``bridge"  by  \citet{lacy07}. In panel (c) we identify the MIR sources, OKYM 2 and 5 from \citet{oka01} and IRS~2E and 2W from \citet{neug69}. In panel (d) we indicate the new infrared counterpart to the mm source
\citep{zap08} detected only at 24.5 \micron\  in the present work (see text).}
\label{irs2mir}
\end{figure}

We present in Figure \ref{irs2mir} the image of IRS~2, which corresponds to the P$_{1}$ pointing in Figure \ref{mosaic}. Each panel of this figure corresponds to an image of IRS~2 taken with one of the filters presented in Table \ref{journal} showing its different aspects. We used the image taken at 12.33 \micron\ to identify the sources, since this is the image with the best S/N ratio.

The overall aspect of IRS~2 in the MIR resembles the ``peanut shaped"\ cloud described by \cite{gww94} in the NIR. According to these authors, at these wavelengths the extended emission is thermal emission from ionized gas associated with the HII regions in W51. At longer wavelengths, the extended emission arises from dust mixed with the ionized gas and heated by trapped Lyman alpha radiation. The five sources identified by \cite{oka01} were detected in our image. However, it is clear from our images that source OKYM~4 is actually an extended source, labeled as ``bridge"  by \cite{lacy07}. This structure is seen also at NIR \citep{lys08} and radio \citep{wc89a} wavelengths (see below).  We confirm that source OKYM 6 is just a ridge of emission extending to the south of IRS~2, extending to the direction of the UCHII region W51d$_{2}$. This fact was first mentioned by \cite{lys08} using the same image at 12.33 \micron\  presented in this work, but it is clearer at longer wavelengths, such as 24.56 \micron.

IRS~2E is the dominant source in all wavelengths, except at 9.69 \micron\ as shown in Figure \ref{irs2mir} (b). This indicates that this source is subject to an extinction much more severe that the other sources. This was noted by Okamoto and collaborators who concluded that IRS~2E (OKYM 1) is further behind the associated HII region and still heavily embedded in the molecular cocoon on that side; this line of sight results in a much higher than average extinction than in general for IRS~2 (see further discussion in the next section).

Figure \ref{irs2mir} (d) is the image of IRS~2 obtained at 24.56 \micron, and at this wavelength, we expect the emission from dust at $\sim$200 K heated by Lyman alpha photons trapped in the UCHII regions. At this wavelength one can detect (and spatially resolve) an embedded MYSO in an evolutionary stage earlier than the UCHII phase, like a Hyper Compact HII (HCHII) region. In this image we detected an object $\sim$1\arcsec\ to the West of IRS~2E, which is not detected at shorter wavelengths. This is the position where \cite{zap08} and \cite{zap09} discovered the massive star which is forming through accretion with high mass accretion rate (4 -- 7 $\times$ $10^{-2}$ $M_{\sun}$yr$^{-1}$); we refer to this object as ZHSRMPG in Figure \ref{mircolor}. This object must be even younger than IRS~2E.

In Figure \ref{mircolor} we show a color composite image, built from the MIR images and the radio emission at 2 cm \citep{wc89a} overplotted on it. The MIR and radio contours were matched based on the coordinates of W51d$_{1}$, which are good within 1\arcsec.  The main radio source in this region, W51d, has a double radio peak. The main peak was correlated with an O3 star \citep[as noted above][]{lys08,barb08} and the secondary was found to match the position of two stellar objects (Barbosa et al. in prep.). Moreover, the secondary radio peak coincides with the MIR peak emission (IRS~2W), where these two stellar objects, presumably massive objects, were found. W51d$_{2}$ was not detected in the \cite{wc89a} radio maps.

Although IRS~2E is the brightest MIR source at $\lambda>2 \micron$, it does not show any emission at radio wavelengths. This fact was first noted by \cite{lys08} and, as pointed out by these authors, it is a consequence of its evolutionary phase: IRS~2E is still in the hot core phase. This may be the case of source OKYM~2, also.

The picture that begins to emerge in IRS~2 is that a significant burst of massive star formation has occurred recently and is ongoing with numerous massive stars (and their associated lower mass stars) yet to have pushed out of their natal cocoons. One very massive star has emerged (at the center of W51d), accompanied by buried sources in IRS~2E (seemingly nearly at the point of breaking out), IRS~2W, and only at the very longest wavelengths the source ZHSRMPG. The counterpart to W51d is to the north and ZHSRMPG to the south suggesting sources are deeper in the natal cloud as one moves north to south.

\subsubsection{The extinction toward IRS~2}\label{extinct}

As we flux calibrated the T-ReCS images, we can now obtain the extinction toward IRS~2, applying the procedures explained in section \ref{highres}, on the images. Therefore we can produce a 2D map of the extinction of the region, the map is presented in Figure \ref{av_map}.

The extinction map of IRS~2 shows two remarkable features. The first one is that source IRS~2E is subject to a high extinction, much greater that the average of rest of the field as noted above. This fact was first noted by \cite{oka01}, who derived an $A_{V}=$137.8--180.9 mag depending on the dust model adopted, and from Table \ref{Td} $A_{V}=63$ mag for IRS~2E. The large reddening of IRS~2E suggests that it is embedded in a dense molecular cocoon behind the ionized gas of the cluster.

The second remarkable feature is the fact that the ``bridge"\ MIR structure found by \cite{lacy07} (represented as a white curved line) is coincident with the regions of the lowest $A_{V}$ within IRS~2. As this structure is interpreted as the region where a stream of high-velocity jet of gas (either a fanlike or a precessing jet) emerges from the molecular cloud, it would naturally lessen the local density of the gas, and so, the extinction is lower as shown in the map.

\subsection{How massive are W51~MAIN and NORTH?}\label{mass_clt}

Although the scope of this work does not include the full characterization of the stellar content of W51~NORTH and MAIN, we can use the results from section \ref{sed_sec} and literature to make a crude estimate of the mass content of W51A and IRS~2 to compare that with other clusters. This can be achieved by comparing the number of stars at the upper limit of the IMF of other stellar clusters with the number of the most massive objects detected in W51. This exercise is useful to put W51 in the context of other clusters. We can assume that the sample of the most massive objects in a given cluster is representative of its total mass, since the IMF appears to be universal for stellar clusters \citep[see][for a review]{kro02}. For this purpose, we compare the total mass of stellar clusters by comparing the number of objects more massive than 40 $M_{\sun}$ (which corresponds to an O6 star) in each cluster. The results of such exercise are shown in the Table \ref{mass}, when available we also show the total mass of the cluster derived from its IMF.

Results showed in Table \ref{mass} put W51 among the most massive clusters in the Galaxy. In particular, W51~IRS~2 is a cluster harboring massive objects in very early stages of their formation and the UCHII region W51d may represent the most massive and evolved objet.

\section{Summary and conclusions}\label{summ}

We presented the results of a study aimed to characterize part of the massive star birth region W51. To that purpose we used public data from the $Spitzer$ space telescope obtained at 3.6, 4.5 and 5.8 \micron\ and data from Gemini mid infrared camera T-ReCS at 7.73, 9.69, 12.33 and 24.56 \micron. The conclusions were presented in each subsection above as part of the discussions, and we summarize them as follows.

Based on the IRAC images from the $Spitzer$ space telescope, we identified MIR counterparts for the UCHII regions W51a, b$_{1}$, b$_{2}$, c$_{1}$, e$_{1}$, e$_{2}$, e$_{7}$ and g. However, UCHII regions b, b$_{3}$ and f do not show any MIR counterpart, but it is very likely that source b could be hidden behind a molecular cloud. Source b$_{3}$ may also be embedded in a gas and dust cocoon, but source f does not show any evidence for a MIR counterpart, either point-like or extended.

\begin{figure}
\centerline{{\scalebox{0.8}{\includegraphics{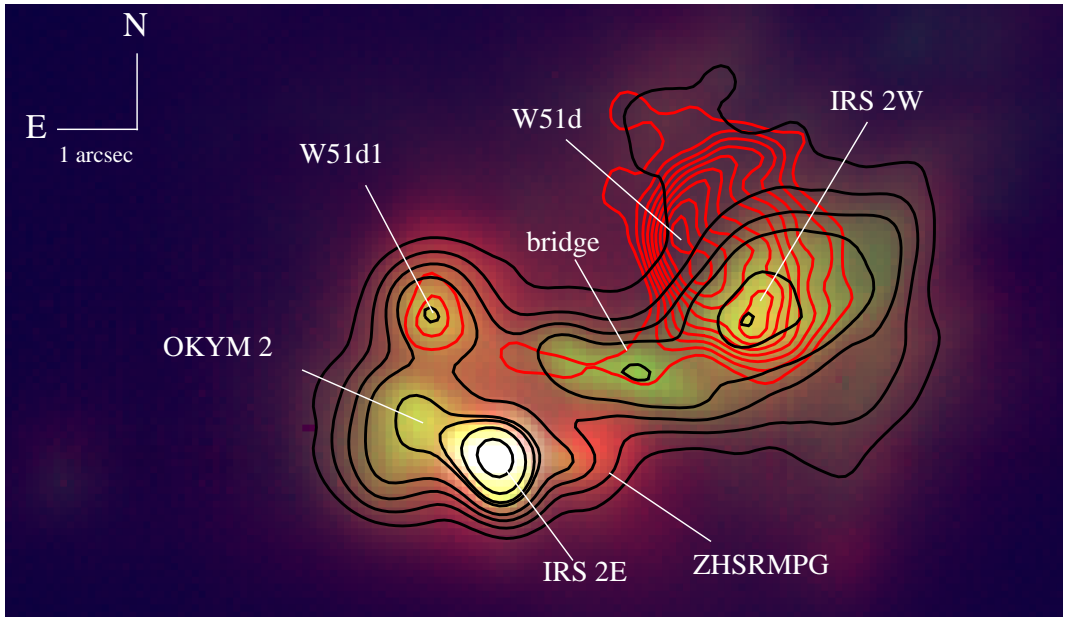}}}}
\caption{Color composite image built from the T-ReCS images coded as: 24.56 \micron\ as the red channel, 12.33 \micron\ as green and 7.69 \micron\ as blue. The black contours represent the emission at 12.33 \micron\ for reference. The red contours represent the radio emission at 2 cm \citep{wc89a}. The level curves are in arbitrary units.}
\label{mircolor}
\end{figure}

\begin{figure}
\centerline{\scalebox{0.6}{\includegraphics{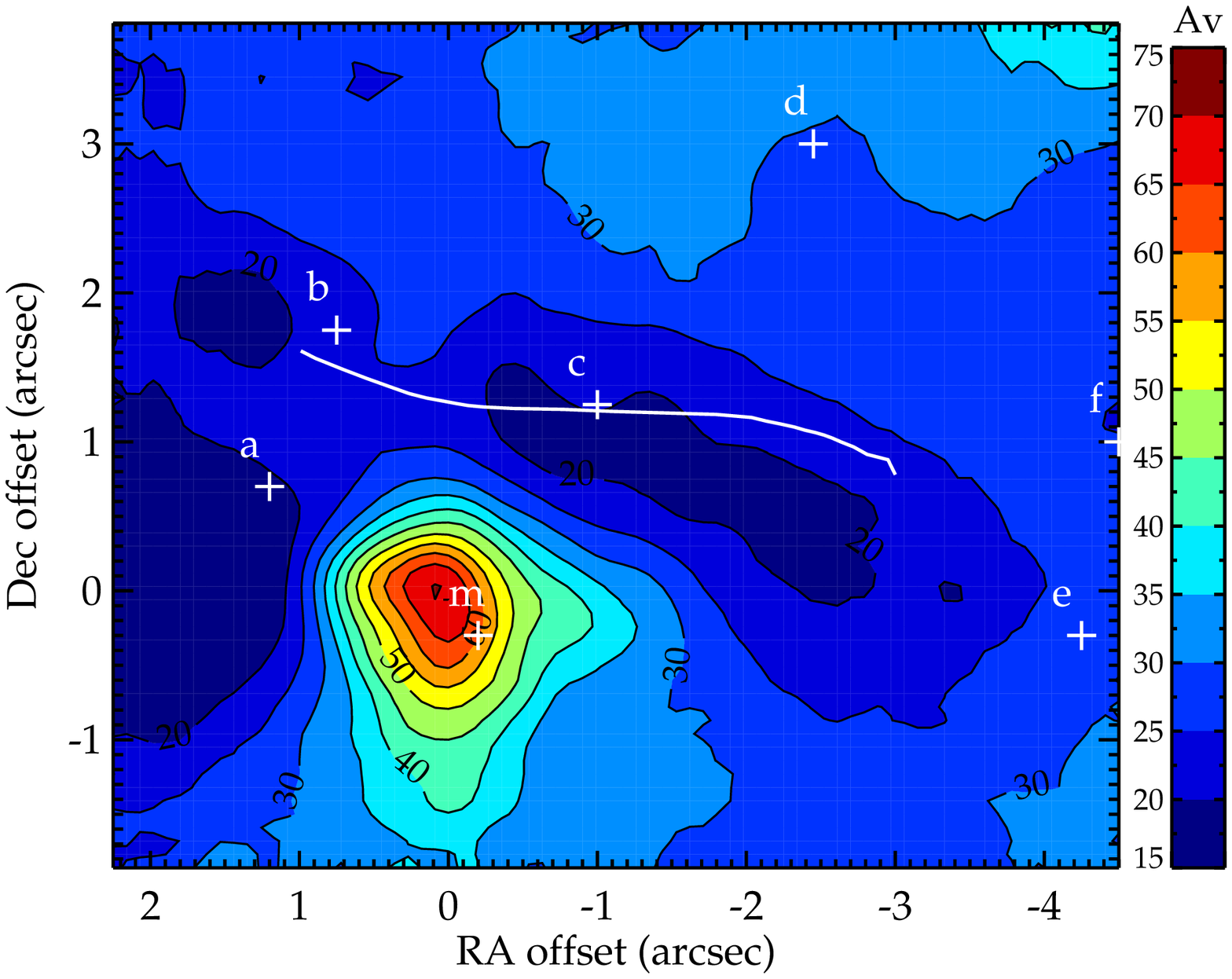}}}
\caption{Extinction map of IRS~2. We adopt the identifications of \citet{lacy07}: sources a-f are HII peaks detected by \citet{meh94}. Comparing to our Figure \ref{mircolor}, source a is the object OKYM~2, source b represents W51d$_{1}$ and d is W51d. The curve represents the region of the blue shifted MIR emission (see text). Source m is the dominant center of the group of SiO masers \citep{sch81,ima02}. Coordinates are centered at the position of IRS~2E, source i on the label scheme of \citet{lacy07}. Both level curves and the color bar at right represents extinction in magnitudes in the $V$ band. The TReCS FOV was cropped to eliminate regions where the S/N ratio was too low.}
\label{av_map}
\end{figure}

Comparing the maps of radio continuum emission obtained at the Very Large Array ($\lambda = 6$ cm), we identified and correlated the radio emission peaks with MIR peaks IRS~1, 2 and 3. IRS~1 has 3 peaks aligned in the N-S direction and the central peak is coincident with the stellar object \#152 presented by \cite{lys08}, which is an extremely red object $(H-K)=5.18$ and, moreover, is associated with UCHII region W51e. IRS~2 is a subcluster of W51 with two MIR peaks IRS~2E and IRS~2W. While the former is a stellar object, the latter is unresolved, but likely includes multiple embedded sources and is centered on a sub-peak of radio emission. Besides these sources, \cite{zap08,zap09} identified a very young massive stellar object at mm wavelengths still in the accretion phase. This object (named ZHSRMPG in Figure \ref{mircolor}) is so embedded in its dust cocoon that it was only detected in wavelength at 25 \micron \ in this work. IRS~3 is a stellar object, already identified in earlier studies, but it is not associated with any radio source.

We modeled MIR point sources with the interactive fitting tool presented by \cite{robi07}. As can be seen in the Table \ref{dat_g1}  the majority of the sources identified using the fitting tool are early B stars as expected. B stars will spend more time than O stars in the embedded phase, so are more likely to be observed. The O types in W51 visible at NIR and MIR wavelengths have already emerged (e.g. W51d).

\cite{oku00} and \cite{kumar04} estimated the age of W51A and IRS~2 in less than 0.7 Myr. Such a timescale is long enough for an early O star to leave the UCHII region phase, which may last $\sim$0.5 Myr \citep{church02} and clean its surroundings. At this phase, the early O stars are not MIR or radio sources anymore, and we would only detect those late O and/or early B stars. Moreover, many of the O stars, still embedded in their UCHII regions are surrounded by intense extended emission making them unobservable at longer wavelengths. The most dramatic case may be IRS~1 which is coincident with UCHII region W51e, which in turn corresponds to the NIR source \#152 from \cite{lys08}. According to the number of ionizing photons derived from radio data of \cite{meh94} and corrected with the calibration tables of \cite{martins05}, its emission is equivalent to the emission of 45 O7V stars, and even so, the sources are still veiled by the extended emission of IRS~1.

The analysis of IRS~1, 2 and 3 allows us to sketch an evolutionary scenario for them. The youngest of all sources is certainly ZHSRMPG, which is still accreting through a high accretion rate that is quenching its radio emission. IRS~2E and IRS~3 may be in an intermediate phase, but still quite young: the MYSO is still surrounded by an accretion disk and may be in the transition from a hot core to an UCHII region \citep{barb08}. IRS~1 is more evolved than the previous sources, as they can be detected in shorter wavelengths, like the NIR $H$ band, and it is a prominent radio source. Finally, W51d is probably the most evolved object among those detected thus far. It is an MYSO still ionizing its UCHII region, cleaned its surroundings sufficiently to allow the detection of photospheric features that reveal it as an O3--type star.

We modeled SEDs of 13 sources detected in the MIR, using data from the 2MASS catalogue for the $J$, $H$ and $K$ bands, the $Spitzer$ GLIMPSE database, at 3.6, 4.5 and 5.8 \micron\ and those obtained in this study at 7.73, 9.69, 12.33 and 24.56 \micron. The results, which may indicate lower limits, show stellar masses for the MYSOs ranging from 4 -- 41 M$_{\sun}$. and were used to make a crude comparison of the cluster mass with other Galactic stellar clusters. Even with the limitations of this procedure, as quoted in the section \ref{mass_clt}, we found that W51 is among the most massive and youngest clusters of the Galaxy.

Using the data obtained with the Gemini MIR camera T-ReCS, we derived the line-of-sight extinction to five sources, ranging from 30 -- 86 mag at the $V$ band. We also calculated the color temperature of the circumstellar dust for 7 sources. The temperatures range from 96 to 140 K, which correspond to the thermal emission from warm dust located at a distance of 0.1 pc from the sources.

Finally we presented the first extinction map of IRS~2, showing how the extinction varies within this region. The ``bridge" structure, interpreted as the region where an underlying jet of gas disrupts a foreground cloud of gas is coincident with a region of lower extinction. We also show that this structure can be detected at shorter wavelengths, such as 2 \micron, to longer wavelengths such as 6 cm.

\acknowledgements

We would like to express our gratitude to Ed Churchwell who kindly made the entire radio catalog of UCHII regions available to us in its electronic version, and to the anonymous referee whose comments improved the quality of the paper.

CLB, and DMG want to thank FAPESP, CNPq and CAPES for financial support. AD thanks FAPESP for continuing financial support. PSC appreciates NSF for continuous financial support. RDB appreciates research support from NOAO/NSF which allowed him to contribute to this work and travel to Brazil to work with CLB and AD. 

Based (in part) on observations obtained at the Gemini Observatory, which is operated by the Association of Universities for Research in Astronomy, Inc., under a cooperative agreement with the NSF on behalf of the Gemini partnership: the National Science Foundation (United States), the Science and Technology Facilities Council (United Kingdom), the National Research Council (Canada), CONICYT (Chile), the Australian Research Council (Australia), Minist\'erio da Ci\^encia, Tecnologia e Inova\c c\~ao (Brazil) and Ministerio de Ciencia, Tecnolog\'{\i}a e Innovaci\'on Productiva (Argentina).

This work is based in part on observations made with the Spitzer Space Telescope, which is operated by the Jet Propulsion Laboratory, California Institute of Technology under a contract with NASA.



{\it Facilities:} \facility{Gemini (T-ReCS)}, \facility{Spitzer (IRAC)}.

\bibliographystyle{apj}
\bibliography{references}

\clearpage

\clearpage

\clearpage

\clearpage

\clearpage

\clearpage

\clearpage

\clearpage

\clearpage

\clearpage

\begin{table}
\caption{Technical details of the observations. The first column lists the filter names. The second column lists the central wavelength of the filter. The third column lists the width (at 50\% transmission) of the filter. The fourth and fifth  columns list the fluxes of the standard stars $\alpha$ Her and $\alpha$ Aql respectively, see text.}

\begin{tabular}{lcccc}
\hline
Filter & $\lambda_{c}$ & $\Delta(\lambda)$ & $\alpha$ Her$^{1}$ & $\alpha$ Aql$^{2}$\\
name & (\micron) & (\micron) & (Jy) & (Jy)\\
\hline
Si-1 & 7.73 & 0.69 & 2011 ($\pm 201$) & 53.2 ($\pm 5.3$)\\
Si-3 & 9.69 & 0.93 & 1460 ($\pm 146 $) & 34.6 ($\pm 3.5$)\\
Si-6 & 12.33 & 1.18 & 1200 ($\pm 120$) & 21.7 ($\pm 2.2$)\\
Qb & 24.56 & 1.92 & 335 ($\pm$) & 5.2 ($\pm$ 0.5)\\
\hline
\label{journal}
\end{tabular}
\\
(1) Fluxes as quoted at \\
http://www.jach.hawaii.edu/UKIRT/astronomy/calib/phot\_cal/bright\_stds.html\\
(2) Fluxes as quoted at\\
http://www.gemini.edu/sciops/instruments/mir/MIRStdFluxes.txt\\
\end{table}

\clearpage

\begin{table}
\caption{Fluxes extracted for each point source detected in the mosaics of Figures \ref{trecs1} and \ref{trecs2}.}

\begin{tabular}{lcccc}
\hline
Source & $F_{7.73}$ & $F_{9.69}$ & $F_{12.33}$ & $F_{24.56}$\\
name & (mJy) & (mJy) & (mJy) & (mJy)\\
\hline
IRS~1/\#1 & 71.5 ($\pm 7$) & 22.4 ($\pm 2$) & 1670 ($\pm 167$) & 19627 ($\pm 196$)\\
IRS~1/\#2 & 344.1 ($\pm 34$) & 13.9 ($\pm 2$) & 641.6 ($\pm 64$) & 12748 ($\pm 127$)\\
IRS~2/\#1 & 36.7 ($\pm 3.7$) & 101.7 ($\pm 10.2$) & 460.7 ($\pm 46.1$) & 12967 ($\pm 129.7$)\\
IRS~2/\#2 & 39.0 ($\pm 4$) & 11.2 ($\pm 1$) & 113.0 ($\pm 11$) & 3676 ($\pm 367$)\\
IRS~2/\#3 & 36.3 ($\pm 4$) & 19.2 ($\pm 2$) & 184.8 ($\pm 18$) & 8674 ($\pm 867$)\\
IRS~2/\#4 & 51.7 ($\pm 5$) & 6.5 ($\pm 0.6$) & 144.3 ($\pm 14$) & 7285 ($\pm 728$)\\
IRS~2E & 10289 ($\pm 1030$) & 330 ($\pm 33$) & 12280 ($\pm 1230$) & 97500 ($\pm 9750$)\\
IRS~3 & 1230 ($\pm 120$) & 12.6 ($\pm 1.3$) & 1785 ($\pm 180$) & 12820 ($\pm 130$)\\
OKYM~2 & 9561 ($\pm 956$) & 2030 ($\pm 203$) & 3960 ($\pm 400$) & $\cdots$\\
W51d$_{1}$ & 400 ($\pm$40) & 880 ($\pm 90$) & 5200 ($\pm 520$) & 48000 ($\pm 4800$)\\
\hline
\label{Av}
\end{tabular}
\end{table}

\clearpage

\begin{table}
\caption{Dust color temperature, dust absorption and visual extinction for each point source detected in the mosaics of Figures \ref{trecs1} and \ref{trecs2}.}

\begin{tabular}{lccc}
\hline
Source & $T_{d}$ & $A_{9.8}$ & $A^{1}_{V}$\\
name & (K) & (mag) & (mag)\\
\hline
IRS~1/\#1 & 128 & 3.8 & 63\\
IRS~1/\#2 & 115 & 3.8 & 63\\
IRS~2/\#1 & 107 & $\cdots$ & $\cdots$\\
IRS~2/\#2 & 104 & 2.0 & 33 \\
IRS~2/\#3 & 98 & 1.8 & 30\\
IRS~2/\#4 & 96 & 2.9 & 48\\
IRS~2E & 140 & 3.8 & 63\\
IRS~3 & 144 & 5.2 & 86\\
OKYM~2 & $\cdots$ & 1.4 & 26\\
W51d$_{1}$ & 135 & $\cdots$ & $\cdots$\\
\hline
\label{Td}
\end{tabular}
\\
(1) The main source of uncertainty is from the adopted extinction law. Assuming \cite{mat90}, for example, $A_{V}\approx18.5\times A_{10 \micron}$
\end{table}

\clearpage

\begin{table}
\caption{Data derived from the best fit models for the SEDs presented in the Figure \ref{sed_g1}.}
\begin{tabular}{lcccc}
\hline
Source & Age$^{1}$ & Mass$^{2}$ & Spectral & Mass Range$^{4}$\\
 & (yr) & (M$_{\sun}$) & Type$^{3}$ & (M$_{\sun}$)\\
 \hline
W51a & 6.84$\times$10$^{4}$ & 5 & B3 & 4.54 -- 6.78\\
W51b$_{1}$ & 2.29$\times$10$^{5}$ & 8 & B1 & 1.53 -- 8.64\\
W51e$_{7}$ & 9.81$\times$10$^{5}$ & 13 & B0.5 & 11.24 -- 16.09\\
IRS~3 & 7.73$\times$10$^{4}$ & 25 & O8 & 10.44 -- 28.43\\
IRS~2/\#1 & 5.33$\times$10$^{4}$ & 9 &  B1 & 7.42 -- 11.51\\
IRS~2/\#2 & 2.97$\times$10$^{5}$ & 7 & B1 & 6.44 -- 9.79\\
IRS~2/\#3 & 1.92$\times$10$^{5}$ & 7 & B1 & 7.31 -- 9.65\\
IRS~2/\#4 & 3.80$\times$10$^{4}$ & 9 & B1 & 5.02 -- 11.33\\
IRS~1/\#1 & 1.56$\times$10$^{5}$ & 18 & B0 & 7.79 -- 18.40\\
IRS~1/\#2 & 3.45$\times$10$^{5}$ & 18 & B0 & 8.96 -- 18.40\\
OKYM~2 & 2.98$\times$10$^{3}$ & 41 & O5.5 & 21.60 -- 41.20\\
W51d$_{1}$ & 1.08$\times$10$^{5}$ & 11 & B0.5 & 7.39 -- 11.07\\
IRS~2E & 1.25$\times$10$^{3}$ & 28 & O7.5 & 13.02 -- 32.64\\
\hline
\label{dat_g1}
\end{tabular}
\\
(1) Best fit age, obtained through an iterative process using evolutionary tracks of \cite{bern96}. See text for details.\\
(2) Best fit mass.\\
(3) Spectral type obtained comparing the stellar mass from the best fit model to those quoted for zero-age main sequence stars by \citet{blum00}.\\
(4) Range of masses of the best 10 fitted models.
\end{table}

\clearpage

\clearpage

\begin{table}
\caption{Spectral types obtained in this work compared to those available in the literature.}
\begin{tabular}{lcc}
\hline
Source & Sp. Type & Sp. Type\\
 & (this work) & (literature)\\
 \hline
W51a & B3 & O6$^{(1)}$\\
W51b$_{1}$ & B1 & O8.5$^{(1)}$\\
W51d$_{1}$ & B0.5 & O8$^{(2)}$\\
W51e$_{7}$ & B0.5 & B1$^{(1)}$\\
\hline
\label{sp_comp}
\end{tabular}
\\
(1) \citet{meh94}, after reviewing the spectral type according to the ionizing fluxes quoted in \citet{martins05}\\
(2) \citet{wc89a}, after reviewing the spectral type according to the ionizing fluxes quoted in \citet{martins05}
\end{table}

\clearpage

\begin{table}
\caption{Comparison of the number of objects more massive than 40$M_{\sun}$ detected in W51~NORTH and MAIN and in other stellar clusters.}
\begin{tabular}{lccl}
\hline
Cluster & Number & Total Mass & Reference\\
Name & of objects & $\times10^{3}(M_{\sun})$ & \\
\hline
Westerlund 1 & $>$50 & $\leq$100 & \cite{neg10}\\
W49 & 72 & 50 - 70 & \cite{hom05}\\
Arches & 25 & 40 & \cite{kim06}\\
NGC 3603 & 14 & 40 & \cite{pang13}\\
Danks 1 & 05 & 8 & \cite{dav12}\\
W51 Reg. 3$^{1}$ & 25 & 8.2 & \cite{oku00}\\
NGC~3576 & 01 & 5.4 & \cite{lys02}\\
G333.1-0.4 & 02 & 1.3 & \cite{lys05}\\
W51 MAIN/NORTH & 02 & \nodata & this work/\cite{barb08}\\
W43 & 29 & \nodata & \cite{blum99}\\
W31 & 04 & \nodata & \cite{blum01}\\
W42 & 01 &  \nodata& \cite{blum00}\\
\hline
\end{tabular}
\label{mass}
\\
(1) Region 3 of \cite{oku00} includes W51~MAIN, NORTH and the UCHII regions W51a, b, b$_{1}$, b$_{2}$, b$_{3}$ and c$_{1}$.
\end{table}

\end{document}